\documentclass[aps,prl,twocolumn,showpacs,superscriptaddress,groupedaddress]{revtex4}  
\usepackage{graphicx}
\usepackage{amsmath}
\usepackage{dcolumn}
\usepackage{bm}

\begin{document}

\title{Magnetic field induced global paramagnetic response in Fulde-Ferrell superconducting strip}

\author{P. M. Marychev}

\affiliation{Institute for Physics of Microstructures, Russian
Academy of Sciences, 603950, Nizhny Novgorod, GSP-105, Russia}

\author{V. D. Plastovets}

\affiliation{Institute for Physics of Microstructures, Russian
Academy of Sciences, 603950, Nizhny Novgorod, GSP-105, Russia}

\affiliation{Lobachevsky State University of Nizhny Novgorod,
Nizhny Novgorod, 603950 Russia}

\affiliation{Sirius University of Science and Technology, 1
Olympic Ave, 354340 Sochi, Russia}

\author{D. Yu. Vodolazov}

\affiliation{Institute for Physics of Microstructures, Russian
Academy of Sciences, 603950, Nizhny Novgorod, GSP-105, Russia}

\begin{abstract}
We theoretically study magnetic response of a
superconductor/ferromagnet/normal-metal (SFN) strip in an in-plane
Fulde--Ferrell (FF) state. We show that unlike to ordinary
superconducting strip the FF strip can be switched from
diamagnetic to paramagnetic and then back to diamagnetic state by
{\it increasing} the perpendicular magnetic field. Being in
paramagnetic state FF strip exhibits magnetic field driven second
order phase transition from FF state to the ordinary state without
spatial modulation along the strip. We argue that the global
paramagnetic response is connected with peculiar dependence of
sheet superconducting current density on supervelocity in FF state
and it exists in nonlinear regime.
\end{abstract}

\maketitle

\section{Introduction}

The diamagnetic Meissner effect, together with zero resistivity,
is the fundamental property of superconducting state. When one
places a superconducting specimen in a weak magnetic field,
screening supercurrents expel magnetic flux from the interior of
superconductor that leads to its diamagnetic response. However,
there are experimental observations of so called paramagnetic
Meissner effect (PME) in high-T$_c$ superconductors
\cite{Svelindh-1989,Riedling-1994} and disks of conventional
superconductors \cite{Thompson-1995,Geim-1998}. But in all these
cases anomalous paramagnetic response was observed only upon
cooling in low magnetic fields and was absent upon cooling without
applied field. For granular high-T$_c$ superconductors the PME can
be explained by the presence of the $\pi$-junctions
\cite{Sigrist-1995}, while in the other cases the PME is caused by
the trapped flux on intrinsic inhomogeneities or surface
\cite{Koshelev-1995,Moshchalkov-1997}.

Paramagnetic response without the captured flux (vortices) can be
realized in case of unusual Cooper pairing, namely the
odd-frequency superconductivity. Odd-frequency pairs formally have
negative density that leads to paramagnetic supercurrents and,
consequently, local paramagnetism \cite{Bergeret-2001}.
Odd-frequency superconducting state can be realized in ferromagnet
part of hybrid superconductor/ferromagnet (SF)
structures\cite{Alidoust-2014}, near the normal metal/p-wave
superconductor (NS) interfaces \cite{Asano-2011} and near the
surface of d-wave superconductors \cite{Walter-1998}. Local
paramagnetic response of odd-frequency superconductivity was
directly observed in SFN trilayer \cite{Bernardo-2015} via
measurement of enhanced magnetic field in normal layer. Also
paramagnetic response of normal metal was seen at ultra-low
temperatures in hybrid superconductor/normal metal structure
\cite{Mota-2000} which could be explained by presence of dilute
magnetic impurities leading to odd-frequency superconductivity
\cite{Espedal-2016}.

In relatively thin SF or SFN strips the paramagnetic response of
odd-frequency superconducting correlations in F or FN layers may
exceed the diamagnetic response of S layer (at proper choice of
material parameters) and the in-plane
Fulde-Ferrell-Larkin-Ovchinnikov (FFLO) state could be developed
\cite{Mironov-2012,Mironov-2018}. It is modulated along the strip
superconducting state and originally its existence was predicted
for a bulk superconductors with spatially uniform exchange field
and energy splitting of electrons with opposite spin of order of
superconducting gap \cite{Fulde-1964,Larkin-1964}. In the FF state
the superconducting order parameter has the form of the plane wave
($\propto exp(i{\bf q}_{FF}{\bf r})$) while in the LO state it is
the standing wave ($\propto cos({\bf q}_{LO}{\bf r})$ near
$T^{FFLO}$). In the pioneer works \cite{Fulde-1964,Larkin-1964} it
was shown that the system being in the FF or LO states retains the
conventional diamagnetic Meissner response at small magnetic
fields.

Here we theoretically show, that magnetic response of SFN strip
being in in-plane Fulde-Ferrell state is also diamagnetic at small
and large fields, but there is finite range of fields where
magnetic response is globally paramagnetic. It differs from global
paramagnetic response predicted for small size unconventional
superconducting disks \cite{Suzuki-2014} and thin disks/squares
made of SFN trilayer \cite{Plastovets-2020}, where it appears due
to finite size effect and exists only at small fields. We argue
that in case of SFN strip global paramagnetic response is
connected with peculiar dependence of sheet superconducting
current density on supervelocity in FF state and it appears in
nonlinear regime (when dependence of superconducting current on
vector potential is nonlinear). The paramagnetic response is
accompanied by magnetic field driven second order phase transition
from FF like state to ordinary state without spatial modulation
along the strip. We also find that in presence of parallel
magnetic field magnetization curves could be different depending
on direction of ${\bf q}_{FF}$ along the strip, which allows one
to determine its direction from magnetic measurements.

\section{Model}

\begin{figure}[hbt]
\includegraphics[width=0.5\textwidth]{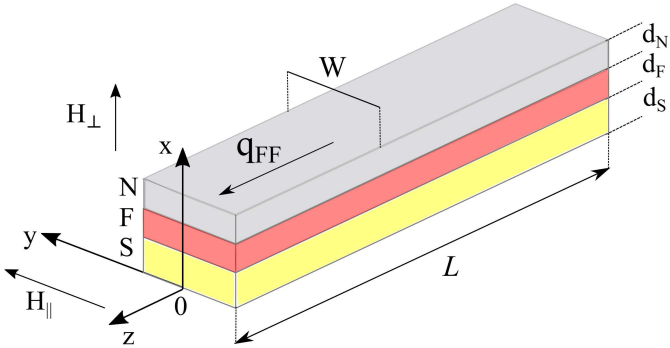}
\caption{\label{Fig:Sys} The schematic representation of the SFN
strip placed in parallel and perpendicular magnetic field.}
\end{figure}

We study magnetic response of SFN strip with length $L$ and width
$w$ made of superconductor with thickness $d_S$, ferromagnet with
thickness $d_F$ and normal metal with thickness $d_N$ (see Fig.
\ref{Fig:Sys}). In Ref. \cite{Mironov-2018} it was shown that when
the ratio of resistivities $\rho_S/\rho_N \gg 1$, thicknesses of S
and N layers are about of coherence length in superconductor and
thickness of F layer is about of coherence length in ferromagnet,
the in-plane Fulde-Ferrell-Larkin-Ovchinnikov state could be
realized (in realistic SF hybrid this state is hard to have due to
large resistivity of F layer). In our work we consider only
Fulde-Ferrell like state because for studied system LO state has
larger energy \cite{Marychev-2018}. In bulk superconductors with
spatially uniform exchange field (magnetic superconductor) LO
state has smaller energy as it was found in Ref.
\cite{Larkin-1964}. This difference could be connected with
properties of SFN trilayer, where superconducting and
ferromagnetic films are thin, spatially separated and there is
gradient of superconducting characteristics across the thickness
of trilayer. It brings difference even between properties of
Fulde-Ferrell states in SFN trilayer and magnetic superconductor.
In both systems in the ground state there is finite phase gradient
$\nabla \varphi= {\bf q}_{FF}$ but in SFN structure there are
finite superconducting currents flowing in S and FN layers in
opposite directions \cite{Mironov-2018} with the total (thickness
integrated) zero current while in magnetic superconductor there is
no spatially separated currents and both local and total currents
are equal to zero.

To calculate the magnetization curve of SFN strip we use two
models. First, we use 2D Usadel equation for normal $g=cos\Theta$
and anomalous $f=sin\Theta\exp(i\varphi)$ quasi-classical Green
functions \cite{Golubov_2004,Buzdin_2005,Bergret_2005}, assuming
that $\Theta$ depends only on x and y and neglect their dependence
on z coordinate
 \begin{equation}
 \label{usadel}
  \begin{split}
\frac{\hbar D}{2}\left(\frac{\partial^2\Theta}{\partial
x^2}+\frac{\partial^2\Theta}{\partial
y^2}\right) \\
-\left((\hbar\omega_n+iE_{ex})+\hbar\frac{D}{2}q^2\cos
\Theta\right)\sin\Theta+\Delta\cos\Theta=0,
\end{split}
 \end{equation}

Here $D$ is the diffusion coefficient of corresponding layer,
$E_{ex}$ is the exchange field which is nonzero only in F layer,
$\Delta$ is the superconducting order parameter which is nonzero
only in S layer, $\hbar \omega_n = \pi k_BT(2n+1)$ are the
Matsubara frequencies ($n$ is an integer number), $q=\nabla\varphi
+ 2\pi\,{\bf A}/\Phi_0$ is the gauge invariant phase gradient that
is proportional to supervelocity $v_s \sim q$ (in this model it
has only z component - see Fig. \ref{Fig:Sys}), $\varphi$ is the
phase of the order parameter, ${\bf A}$ is the vector potential,
$\Phi_0=\pi\hbar c/|e|$ is the magnetic flux quantum. $\Delta$
should satisfy the self-consistency equation
\begin{equation}
\label{self-cons}
\Delta \ln\left(\frac{T}{T_{c0}}\right)=2\pi k_B
T\sum_{\omega_n
>0} Re\left(\sin\Theta_S - \frac{\Delta}{\hbar\omega_n}\right),
\end{equation}
where $T_{c0}$ is the critical temperature of single S layer in
the absence of magnetic field. Equation (\ref{usadel}) are
supplemented by the Kupriyanov-Lukichev boundary conditions
between layers \cite{JETP-1988}
  \begin{eqnarray}
  \nonumber
    \left.D_S\frac{d\Theta_S}{dx}\right|_{x=d_S-0}=\left.D_F\frac{d\Theta_F}{dx}\right|_{x=d_S+0},
    \\
    \label{boundary}
    \left.D_F\frac{d\Theta_F}{dx}\right|_{x=d_S+d_F-0}=\left.D_N\frac{d\Theta_N}{dx}\right|_{x=d_S+d_F+0}
    \end{eqnarray}

We assume transparent interfaces between layers and thereupon
$\Theta$ is continuous function of x. For interfaces with vacuum
we use the boundary condition $d\Theta/dn=0$.

Because the thickness of whole structure is much smaller
than the London penetration depth $\lambda$ of the single S layer
we neglect the contribution to vector potential from screening currents.
In calculations we use the following vector potential: ${\bf
A}=(0,0,-H_{\parallel}x+H_{\perp}y)$, where $H_{\parallel}$ is the parallel and $H_{\perp}$ is perpendicular magnetic field
(see Fig. \ref{Fig:Sys}).

We calculate the magnetization $\bf{M}$ as
\begin{equation}
 \label{magn}
{\bf M}=\frac{{\bf m}}{dw}=\frac{1}{2cdw}\int\int [{\bf r\times
j_s}] dxdy,
\end{equation}
where ${\bf j}_s=(0,0,j_z)$ is the superconducting current density

\begin{equation}
 \label{current}
j_z(x,y)=\frac{2\pi k_BT}{e\rho}q\sum_{\omega_n > 0}\Re(\sin^2\Theta),
\end{equation}
and we are interested in x component of magnetization $M_x$.

In numerical calculations we use the dimensionless units. The
magnitude of the order parameter is normalized in units of
$k_BT_{c0}$, length is in units of $\xi_c=\sqrt{\hbar
D_S/k_BT_{c0}}$. The magnetic field is measured in units of
$H_s=\Phi_0/2\pi w\xi_c$, magnetization $M_x$ is in units of
$M_0=\Phi_0/2\pi\xi_c^2$. We also take in calculations that
$\lambda(0)/\xi_c=50$, where $\lambda(0)$ is London penetration
depth in single S layer at $T=0$.

To find $j_z$ and $M_x$, we numerically solve equations
(\ref{usadel},\ref{self-cons}) with corresponding boundary
conditions. To reduce the number of free parameters we assume that
the resistivity of S and F layers are equal, i.e.
$\rho_S/\rho_F=1$, which roughly corresponds to parameters of real
highly resistive S and F films. We use $\rho_S/\rho_N=150$ in our
calculations because formation of FF state in the SFN structure
needs the large ratio of resistivities of N layer and S layers
\cite{Mironov-2018}. It corresponds, for example, to pair NbN/Au.

The model above is not able to take into account the states with
dependence of $\Theta$ on longitudinal coordinate (for example
vortex state). To obtain full in-plane distribution of the
superconducting order parameter and current density, one has to
solve 3D Usadel equation which is complicated problem. Instead we
use the Ginzburg-Landau like approach and describe SFN structure
by the 2D (in y and z directions) equations with the effective
superconducting order parameter $\Psi$ averaged over the thickness
of SFN trilayer \cite{Plastovets-2020}. The GL free energy
functional describing 2D superconductor being in the FFLO phase
was proposed in Ref. \cite{Buzdin_2007}
\begin{align}\label{eq:gl_1}
\widetilde{F}=\alpha(T)|\widetilde{\Psi}|^2
+\frac{\beta}{2}|\widetilde{\Psi}|^4+\gamma(|\Pi_y\widetilde{\Psi}|^2+|\Pi_z\widetilde{\Psi}|^2)
\\ \notag
+\delta(|\Pi^2_y\widetilde{\Psi}|^2+|\Pi^2_z\widetilde{\Psi}|^2+|\Pi_y\Pi_z\widetilde{\Psi}|^2
+|\Pi_z\Pi_y\widetilde{\Psi}|^2),
\end{align}
where $\widetilde{\Psi}$ is a complex superconducting order
parameter and $\Pi_{y,z}=\nabla_{y,z}+i2\pi A_{y,z}/\Phi_0$. One
has to define the signs of phenomenological parameters: $\alpha$,
$\gamma<0$ and $ \beta$, $\delta >0$ to have Fulde-Ferrell state
as a ground one \cite{Samokhin_2017,Plastovets_2019}. We have to
stress that for SFN trilayer this functional was not derived from
microscopic theory and we use it as phenomenological one.
\begin{figure}[hbt]
\includegraphics[width=0.48\textwidth]{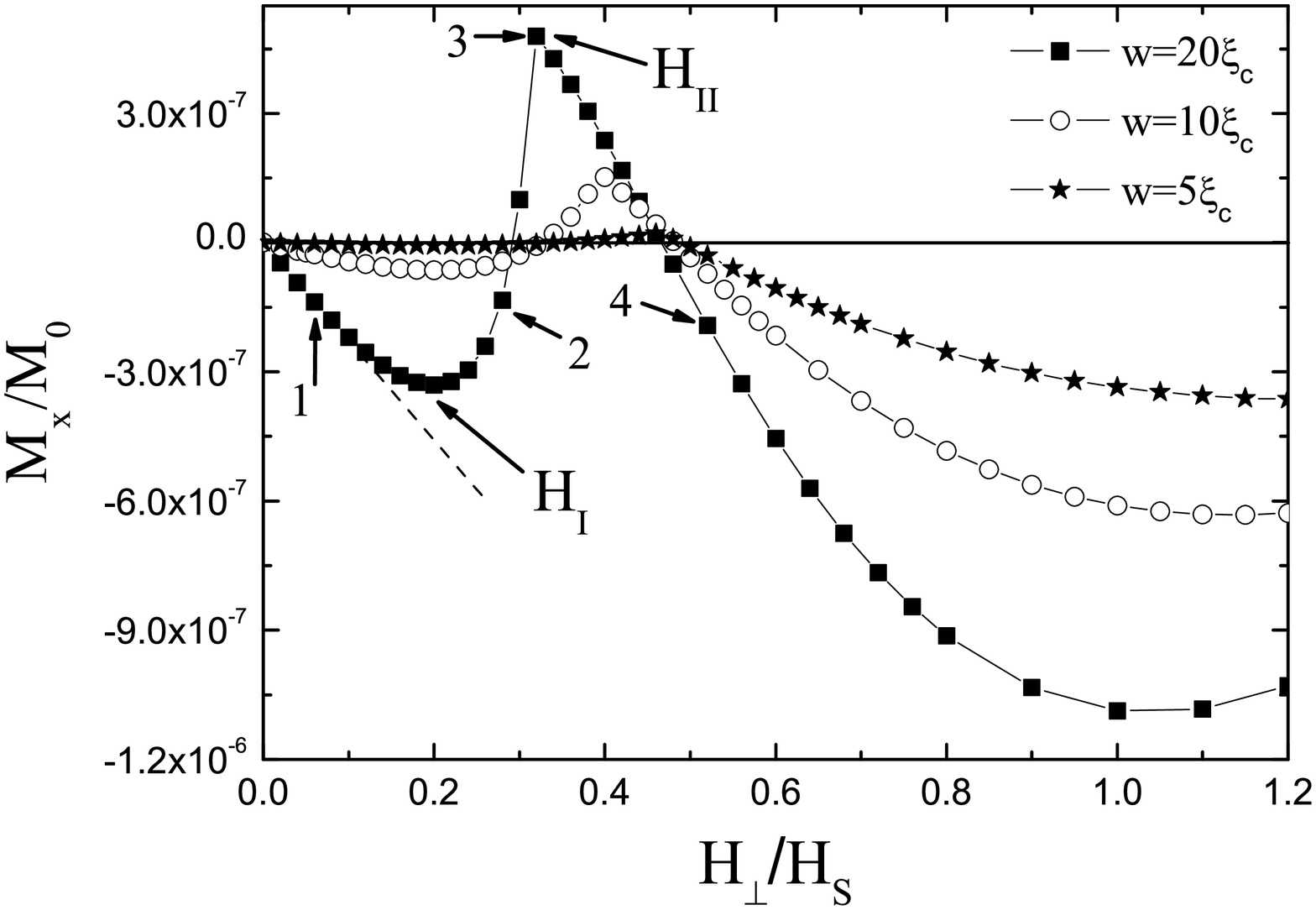}
\caption{\label{Fig:M(h)_w} The magnetization curves of SFN strips
with different widths, found from Usadel model. At $H_{\perp}=H_I$
there is a local minimum in dependence $M_x(H_{\perp})$. At field
$H_{\perp}=H_{II}$ there is second order transition from the state
with $\overline{q_z} \neq 0$ ($H_{\perp}<H_{II}$ - FF like state)
to the state with $\overline{q_z}=0$ ($H_{\perp} \geq H_{II}$ -
ordinary state). Numbers 1-4 indicate fields, at which
distribution of sheet current density over the width of SFN strip
is shown in Fig. 3(a). The parameters of SFN strips are following:
$w=5, 10, 20 \xi_c$, $d_S=1.1\xi_c$, $d_F=0.5\xi_c$, $d_N=\xi_c$,
$E_{ex}=5k_BT_{c0}$ and $T=0.2T_{c0}$.}
\end{figure}

The dimensionless free energy $F$ and order parameter $\Psi$ are
introduced as: $F=F_{GL}\widetilde{F}=(\alpha^2/\beta)
\widetilde{F}$, $\Psi=\Psi_0\widetilde{\Psi}=\sqrt{|\alpha|/\beta}
\widetilde{\Psi}$, with the characteristic length
$\xi_{GL}=\sqrt{|\gamma|/|\alpha|}$ and the dimensionless
parameter $\zeta=|\alpha|\delta/|\beta|^2$. Varying $\int F{\text
dS}$ with respect to $\widetilde{\Psi}^*$ we obtain the
Ginzburg-Landau equation for the dimensionless order parameter:
\begin{gather}\label{eq:gl_2}
\zeta \{\Pi^4_y+\Pi^2_y\Pi^2_z+\Pi^2_z\Pi^2_y+\Pi^4_z\}\Psi
\\ \notag
+\{\Pi^2_y+\Pi^2_z\}\Psi+\Psi|\Psi|^2-\Psi=0.
\end{gather}
Equation (\ref{eq:gl_2}) is supplemented by the boundary
conditions
\begin{gather}\label{eq:gl_3}
\Pi \Psi\Big|_n=0, \quad \Pi^3 \Psi\Big|_n=0.
\end{gather}
which provide vanishing of normal component of superconducting
current $j_s|_n$ and supermomentum $q|_n=(\nabla \varphi+2\pi
A/\Phi_0)|_n$ on the boundary of FF strip with vacuum
\cite{Plastovets-2020}.

In GL model we find $M_x$ by numerical differentiation of
$F_{GL}(H_{\perp})$
\begin{equation}
M_x=-\frac{dF_{GL}}{dH_{\perp}}.
\end{equation}
In principle, the same could be done in Usadel model too, without
using of Eq. (4), but it needs small step in $H_{\perp}$ and very
large calculation time. It is the reason why we use different
methods to find $M_x(H_{\perp})$ in Usadel and GL models.

We use relaxation method with adding of the time derivative
$\partial \Psi/\partial t$ in the right hand side of Eq.
(\ref{eq:gl_2}) and looking for $\Psi(y,z)$ which does not depend
on time. In numerical calculations we put $\zeta =1/8, 1/2, 2, 4$.
Case $\zeta \lesssim 1/2$ corresponds to situation when coherence
length $\xi=\xi_{GL}(2\zeta/((1+4\zeta)^{1/2}-1))^{1/2}$
(characteristic length variation of $|\Psi|$ in used model) is
larger than $q^{-1}_{FF}=\xi_{GL}\sqrt{2\zeta}$ while for $\zeta
\gtrsim 1/2$ we have opposite case, which corresponds to
properties of SFN strip with realistic parameters.

\section{Magnetic response of a SFN strip being in the FF state}

In Fig. \ref{Fig:M(h)_w} we present dependence $M_x(H_{\perp})$,
found in Usadel model, for the SFN strips with different widths
being in FF state at $H_{\perp}=0$. The magnetic response is
diamagnetic at small fields as in ordinary superconductors and
magnetic superconductors with spatially uniform exchange field
\cite{Fulde-1964,Larkin-1964} but at some field (we mark it as
$H_{I}$ in Fig. 2) $M_x$ reaches minimal value and then it becomes
nonmonotonic function of $H_{\perp}$ and changes sign twice. As a
result there is finite range of magnetic fields where magnetic
response is paramagnetic. Moreover, at field $H_{\perp}=H_{II}$
(see Fig. 2) there is a kink, which is a signature of second order
phase transition from the state with $\overline{q_z} \neq 0$
($\overline{q_z}=\int q_z dy/w$ is width averaged $q_z$) to the
state with $\overline{q_z}=0$.
\begin{figure}[hbt]
\includegraphics[width=0.42\textwidth]{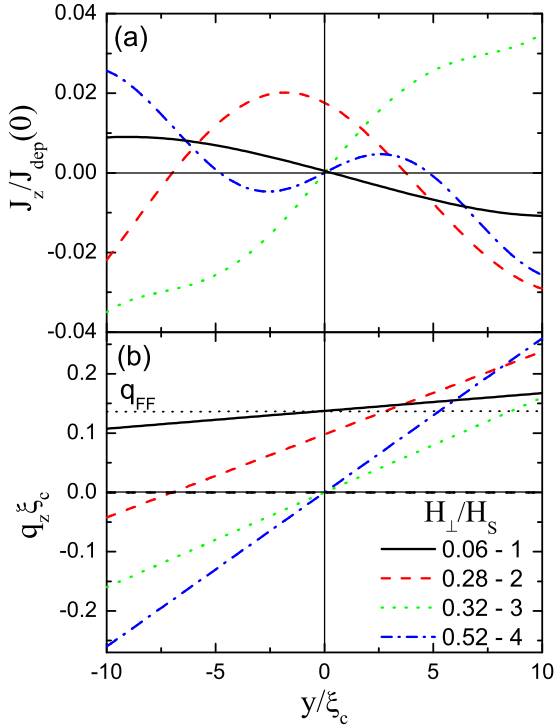}
\caption{(a) Distribution of sheet current density $J_z$ and (b)
supervelocity $\sim q$ over SFN strip with width 20$\xi_c$ at
different $H_{\perp}$ marked by numbers 1-4 in Fig. 2. At
$H_{\perp} = 0.32 H_s$ the $\overline{q} = 0$. $J_z$ is normalized
in units of $J_{dep}(0)=j_{dep}(0)d$, where $j_{dep}(0)$ is
depairing current density of single S layer at $T=0$.}
\end{figure}

To explain this behavior in Fig. 3(a,b) we show distribution of
sheet current density $J_z=\int j_zdx$ and supervelocity $\sim
q_z$ over the width of SFN strip and in Fig. 4 dependence of
$J_z(q_z)$ in spatially homogenous case ($q_z(y)=const$ and
$J_z(y)=const$). When $H_{\perp}=0$ in the ground state of FF
strip there is a finite phase gradient $\nabla \phi=q_{FF}$ but
$J_z(q_{FF})=0$. From Fig. 4 one can see that near $q_z=q_{FF}$
there is London like relation $J_z(y) \sim J_z(q_{FF})+(2\pi
A_z(y)/\Phi_0)dJ_s/dq_z\sim -A_z(y)$ which leads to diamagnetic
response of FF strip at small fields (see Fig. 2). At that fields
dependence $J_z(y)$ is nearly odd function of $y$ ($J_z(y)\sim
-J_z(-y)$ - see Fig. 3(a) for $H_{\perp}=0.06 H_s$) as in ordinary
strip because $dJ_s/dq_z$ is almost constant at $q_z\simeq q_{FF}$
- see dashed line in Fig. 4.

At larger fields due to different nonlinearity of $J_z(q_z)$ at
$q_z<q_{FF}$ and $q_z>q_{FF}$ the width averaged $\overline{q_z}$
($\overline{q_z}(H_{\perp}=0)=q_{FF}$) decreases, as it could be
seen from Fig. 3(b), to provide zero full current $\int J_zdy=0$
and $J_z(y)$ is not odd function of $y$ (see Fig. 3(a) for
$H_{\perp}=0.28 H_s$). As a side effect it leads to nonmonotonous
change of $|M_x|$ and even to paramagnetic response because on
dependence $J_z(q)$ there is a region ($0<q_z<q_{c1}$) where
$dJ_z/dq_z>0$. In current driven regime with $q_z(y)=const$ this
region is not accessible \cite{Samokhin_2017} but it can be
reached, as we find here, with coordinate dependent $q_z(y)$.

The $\overline{q_z}$ goes to zero at $H_{\perp}=H_{II}$ and
simultaneously $M_x$ reaches maximal positive value. $M_x$
decreases and than changes sign at $H_{\perp}>H_{II}$ while
$\overline{q_z}=0$. Therefore at $H_{\perp}=H_{II}$ there is
second order phase transition from the state with $\overline{q_z}
\neq 0$ (Fulde-Ferrell like state) to the state with
$\overline{q_z} = 0$ which is manifested as a kink on dependence
$M_x(H_{\perp})$ (see Fig. 2).
\begin{figure}[hbt]
\includegraphics[width=0.5\textwidth]{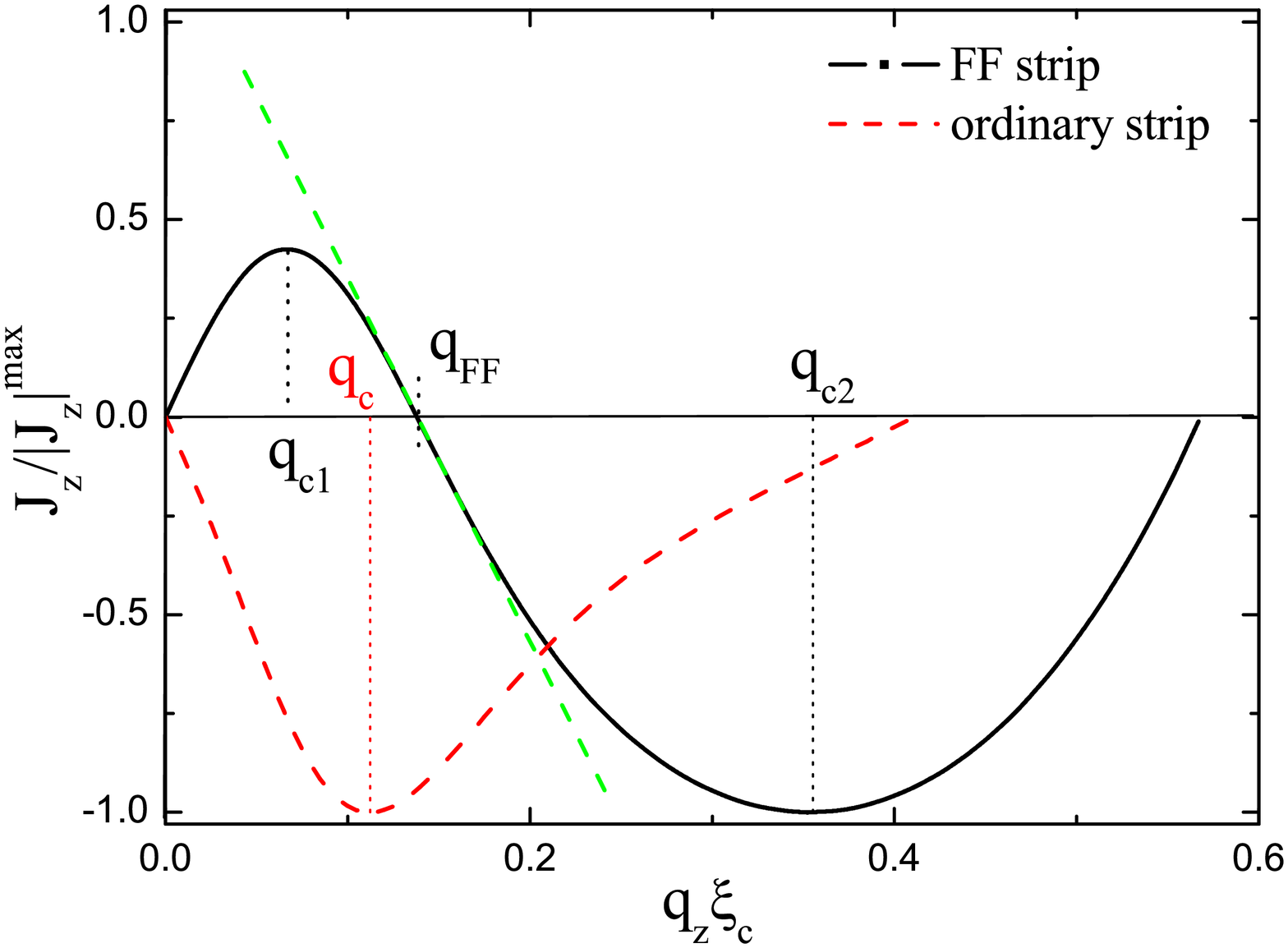}
\caption{Dependence of sheet current density $J_z$ on $q_z$ in
spatially uniform case ($J_z(y)=const$) for SFN strip (parameters
as in Fig. 2) being in FF state and SFN strip (parameters as in
Fig. 2 except $d_F=0.2\xi_c$) being in ordinary state. $J_z$ is
normalized by critical current density (it corresponds to maximal
$|J_z|=|J_z|^{max}$).}
\end{figure}

In ordinary superconducting strip the vorticies enter the sample
when supervelocity at the edge exceeds critical value ($|\pm
q_z(w/2)| \gtrsim q_c$) \cite{Vodolazov_2001}, except rather
narrow strips with $w \lesssim 2 \xi(T)$ which do not have space
for vortex \cite{Fink_1969,Sarma}. We expect similar behavior for
FF strip too and it is the reason why we do not present
$M_x(H_{\perp})$ in Fig. 2 at large fields where $q_z(w/2)$ well
exceeds $q_{c2}$ ($q_z(w/2)=q_{c2}$ at $H_{\perp}=0.71H_s$ for
chosen parameters). But for FF strip we have additional critical
value - $q_{c1}$ (see Fig. 4). Note that $q_z(-w/2)$ becomes
smaller than $q_{c1}$ (it occurs at $H_{\perp}\lesssim H_{I}$)
before $M_x$ changes the sign, which means that the instability
may occur which breaks the homogenous along the strip state and
changes dependence $M_x(H_{\perp})$. To check it we calculate
magnetic response of FF strip of finite length using
Ginzburg-Landau model.

We find that while width of the strip is smaller than $w_c\sim
2q^{-1}_{FF}$ the evolution of $M_x$ and $\overline{q_z}$ with
magnetic field is similar to ones found from Usadel model (compare
Fig. 5(a) and Fig. 2). There is a range of magnetic fields where
the magnetic response is paramagnetic and at $H=H_{II}$ there is
second order transition to state with $\overline{q_z}=0$. At
larger fields magnetic response again becomes diamagnetic and if
the width of the strip is larger than $\sim 2 \xi$ vortices enter
the FF strip which leads to jumps in $M_x$ as in ordinary
superconducting strip (see Fig. 5(a)). Moreover, even relative
change of magnetization is similar in Figs. 2 and 5 (if we
compare, for example, maximal positive and negative $M_x$). Note,
that in Figs. 2 we present results found in Usadel model for SFN
structure with realistic parameters, and it helps to estimate the
strength of the effect (see section Summary). In Fig. 5 we present
results found in GL model, where $M_{GL}$ is some parameter which
we cannot express via material characteristics of SFN structure.

\begin{figure}[h]
\includegraphics[width=0.45\textwidth]{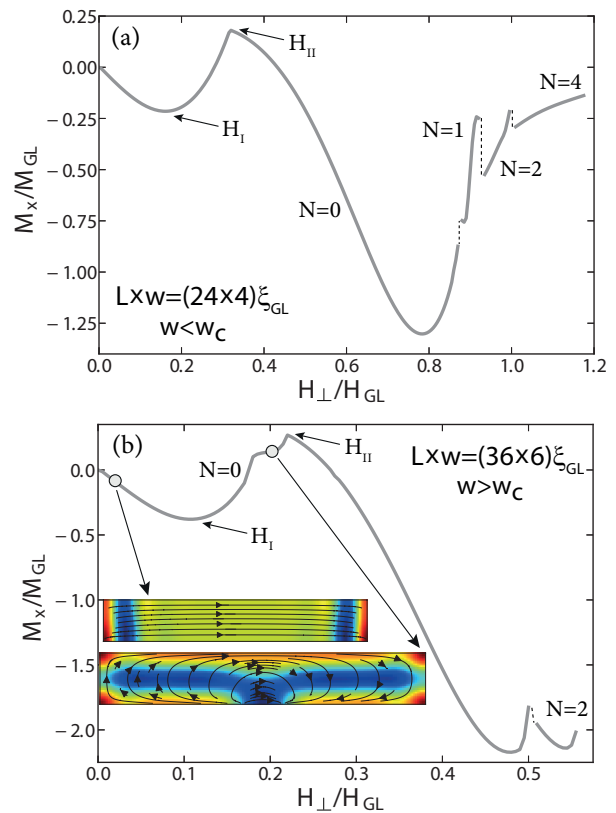}
\caption{\label{Fig:M(h)_GL} Field-dependent magnetization of FF
strips calculated in framework of Ginzburg-Landau model. Lateral
sizes of FF strip are shown in panels (a) and (b), parameter
$\zeta=2$ ($q^{-1}_{FF}=2\xi_{GL} > \xi=\sqrt{2}\xi_{GL}$).
Magnetic field is measured in units of
$H_{GL}=\Phi_0/2\pi\xi^2_{GL}$, magnetic moment is in units of
$M_{GL}=F_{GL}/H_{GL}$, $N=\oint\nabla \varphi \text{d{\bf
l}}/2\pi$ is a total vorticity in the strip. In inset in (b) we
show spatial distribution of $|\Psi|$ and $q$ at different
magnetic fields.}
\end{figure}

For strip with $w>w_c$ the evolution of $M_x$ and $\overline{q_z}$
in field range $H_{I} \lesssim H_{\perp} \leq H_{II}$ is
different. It turns out that at $H_{\perp} \gtrsim H_I$ there
appears finite $q_y$ (transversal component of ${\vec q}$) not
only near the ends of the strip, where it provides conservation of
full current, but also far from it (see insets in Fig. 5(b)). In
different halves of the strip $q_y$ has opposite sign due to
different sign of the screening currents. In regions where $q_y
\neq 0$ $\overline{q_z}$ is got suppressed, depends on
longitudinal (z) coordinate and it has maximum in the center of
the strip. With increasing of magnetic field $\overline{q_z}(z)$
gradually decreases and at $H = H_{II}$ it goes to zero along the
whole strip.

\begin{figure}[h]
\includegraphics[width=0.45\textwidth]{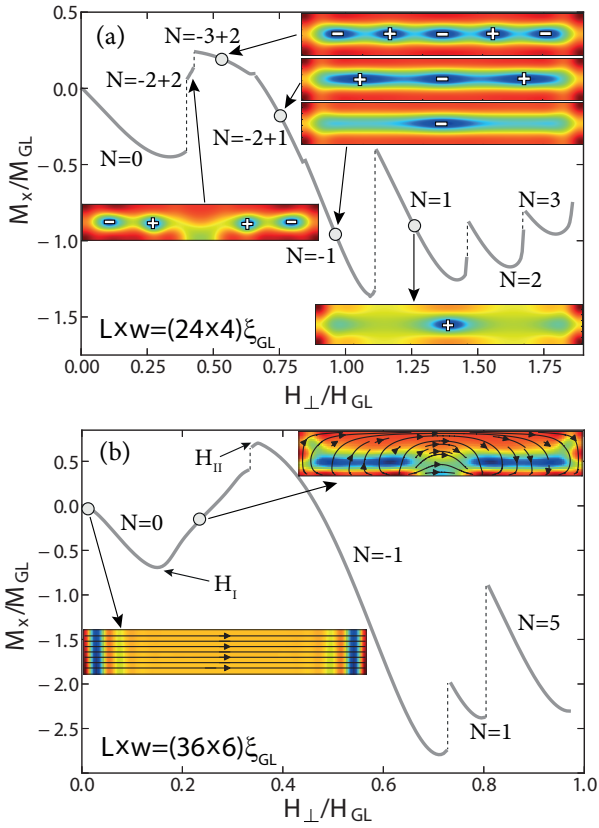}
\caption{\label{Fig:M(h)_GL} Field-dependent magnetization of FF
strips calculated in framework of Ginzburg-Landau model. Lateral
sizes of FF strip are shown in panels (a) and (b), parameter
$\zeta=0.5$ ($q^{-1}_{FF}=\xi_{GL} < \xi \simeq 0.84 \xi_{GL}$).
For both panels $w>w_c$. In insets we show spatial distribution of
$|\Psi|$ and $q$ at different magnetic fields. Symbols $-$ and $+$
indicate antivortex and vortex, correspondingly.}
\end{figure}

Apparently, found critical width of the strip $w_c\sim
2q^{-1}_{FF}$ is correlated with critical length of quasi 1D FF
superconductor $L_c=\pi/\sqrt{2}q^{-1}_{FF}\simeq 2.2 q^{-1}_{FF}$
when spatially modulated state with $q \neq 0$ can appear
\cite{Plastovets-2020}. In narrower strip the transition to state
with $\overline{q_z}=0$ occurs homogenously along the strip
because $\overline{q_z}$ depends on z only near the ends where
$q_z=0$ due to boundary conditions and results found in framework
of Usadel model and GL models qualitatively coincide. In wider
strip $\overline{q_z}$ strongly depends on length at $H_{\perp} >
H_I$ because of appearance of transversal component of ${\vec q}$.
Obviously this result cannot be found in framework of our 2D
Usadel model which assumes spatial uniformity ($q_z(z)=const$)
along the FF strip. It leads to quantitatively different shape of
$M_x(H_{\perp})$ in field range $H_I<H_{\perp}<H_{II}$ for strips
with $w>w_c$ and $w<w_c$ (compare Fig. 5(b) with Fig. 5(a) and
Fig. 2). For parameters of SFN strip those magnetic response is
shown in Fig. 2 $q^{-1}_{FF} \sim 7.2 \xi_c$ (see Fig. 4) and,
hence, only for strip with $w=20 \xi_c$ we can expect appearance
of transversal modulation.

We also find interesting behavior when $\zeta=1/2$ and $1/8$ which
physically correspond to $\xi \gtrsim q^{-1}_{FF}$. In strip with
$w \gtrsim w_c$ the transition to state with $\overline{q_z}=0$
starts from the ends of the strip but it is accompanied not only
by appearance of finite $q_y$ but also vortex-antivortex pairs
(see insets in Fig. 6(a)). In longer strip their number increases
with increasing of magnetic field and reaches the maximal value at
$H = H_{II}$ (for example when $L=48 \xi_{GL}$ there are four
vortex-antivortex pairs - not shown here). At $H=H_{II}$ there is
first order transition to state with $\overline{q_z}=0$ and one
additional antivortex enters the strip in its center (see Fig.
6(a)). With further increase of magnetic field the number of
vortex-antivortex pairs decreases one by one and at large field
only vortices exist in the strip. In wider strip (see Fig. 6(b))
vortex-antivortex pairs do not appear but transition at $H=H_{II}$
is also of first order and one antivortex enters the strip in its
center which is annihilated with vortices at larger fields (see
Fig. 6(b)).

In ordinary superconductors vortices and antivortices can coexist
in small size (mesoscopic) samples placed in external magnetic
field \cite{Chibotaru_2000,Mel'nikov_2002,Misko_2003} or near
ferromagnetic domain wall where magnetic field changes the sign
(experimentally such vortices and antivortices have been observed
recently in ferromagnetic superconductor
EuFe$_2$(As$_{0.79}$P$_{0.21}$)$_2$ \cite{Stolyarov_2018}). In
zero magnetic field their simultaneous appearance in the ground
state was predicted in FFLO system with two coupled
superconducting order parameters \cite{Samoilenka_2020} and as a
metastable state they may exist in small size FF superconductor
\cite{Plastovets-2020}. We find that in FF strip vortex-antivortex
chain is a ground state in finite range of the magnetic fields
when $w \gtrsim w_c$  and $\xi \gtrsim q^{-1}_{FF}$.

\begin{figure}[hbt]
\includegraphics[width=0.45\textwidth]{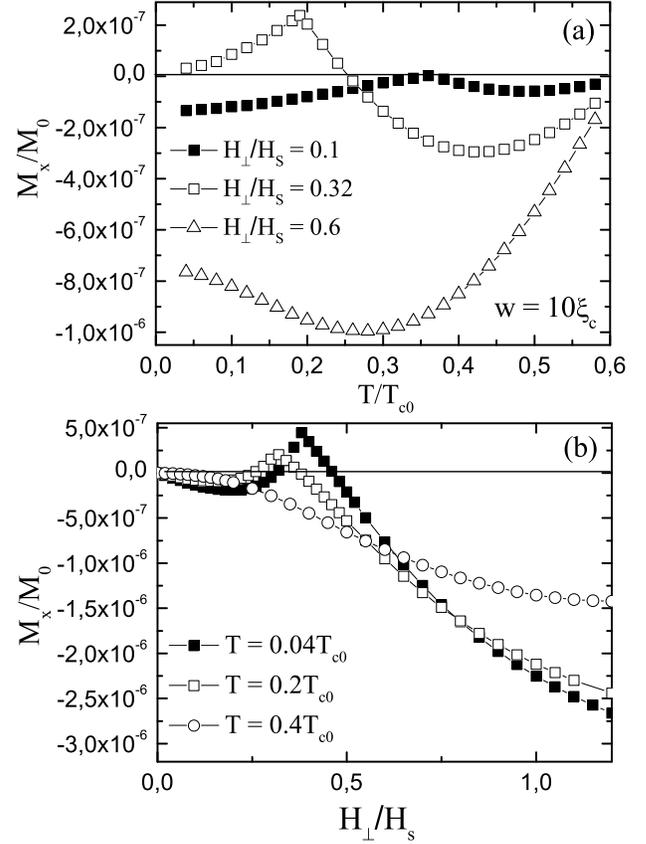}
 \caption{\label{Fig:M(h)_t}
(a) Dependence of the magnetization of SFN strip on the
temperature at different values of the perpendicular magnetic
field. (b) The magnetization curve of the SFN strip at different
temperatures. We use the same parameters of SFN strip as in Fig. 2
except $d_S=1.4 \xi_N$ and choose $w=10\xi_c$. The temperature of
transition to the FF state is $T^{FF}=0.38T_{c0}$, critical
temperature of SFN trilayer is $T_c=0.62 T_{c0}$ (both at
$H_{\perp}=0$).}
\end{figure}

Using Usadel approach we also calculate dependence $M_x(T)$ at
fixed $H_{\perp}$ and $M_x(H_{\perp})$ at different temperatures
(see Fig. \ref{Fig:M(h)_t}). We use the same parameters as in Fig.
2, except thickness of S layer was chosen $d_S=1.4 \xi_c$, for
which the transition temperature to FF state $T^{FF}$ is below the
critical temperature of trilayer $T_c=0.62T_{c0}>T^{FF}=0.38
T_{c0}$. It can be seen that $M_x(T)$ at small fields is
nonmonotonous even at $T>T^{FF}$, which is consequence of
existence of paramagnetic currents in FN layers, while the global
magnetic response is diamagnetic for all fields when $T>T^{FF}$
(see Fig. \ref{Fig:M(h)_t}(b)).

\begin{figure}[hbt]
\includegraphics[width=0.5\textwidth]{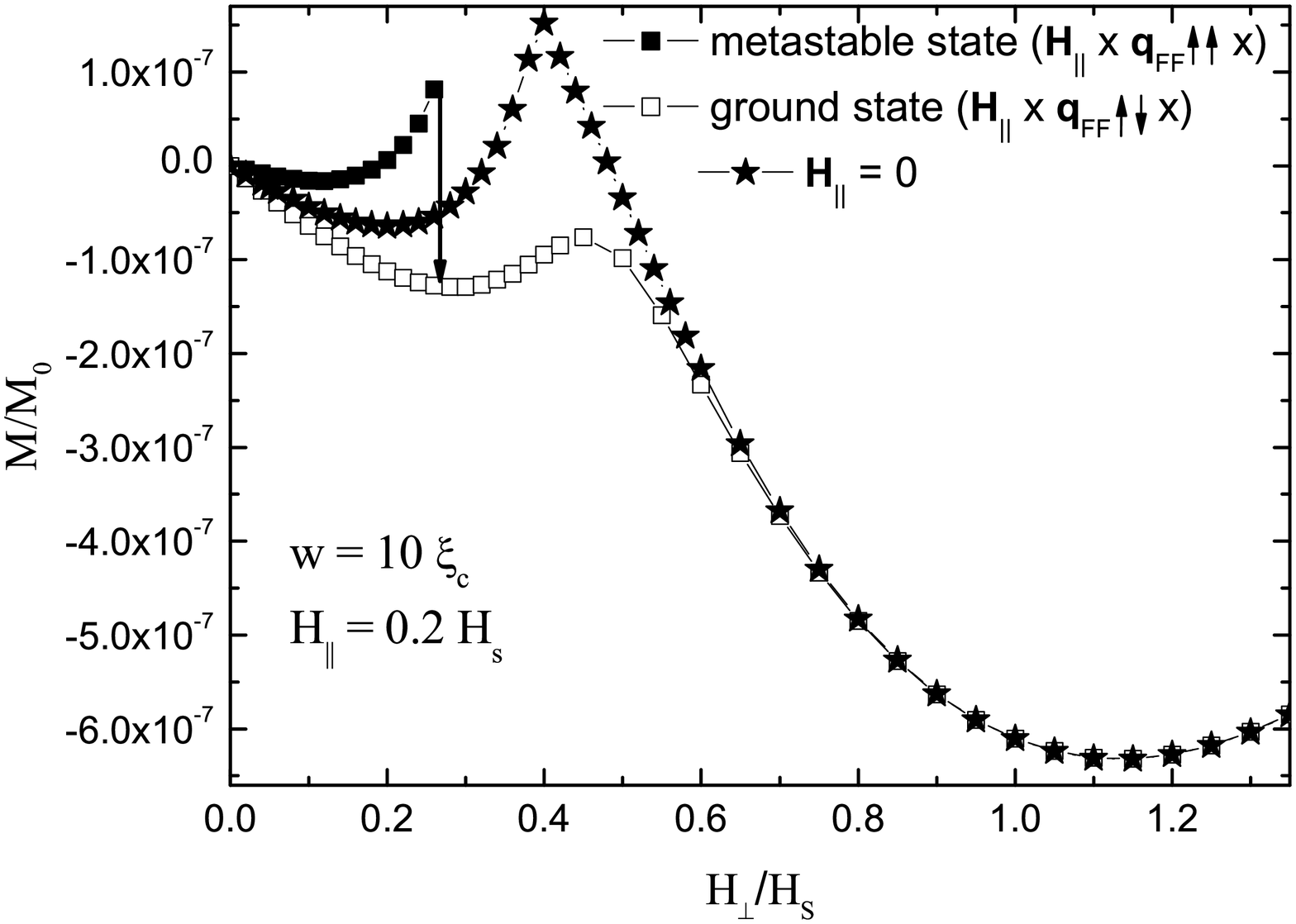}
\caption{\label{Fig:M(h)_par} The magnetization curves of SFN
strip being in ground and metastable FF states (having opposite
${\bf q}_{FF}$) which are controlled by parallel magnetic field.
For comparison we also present $M_{x}(H_{\perp})$ when
$H_{\parallel}=0$. The arrow indicates the direction of
magnetization jump which occurs with increase of $H_{\perp}$.
Width of SFN strip $w=10\xi_c$, $H_{\parallel}=0.2H_s$, the other
parameters are as in Fig. \ref{Fig:M(h)_w}. At
$H_{\parallel}>H_{\parallel}^* \simeq 0.4 H_s$ and $H_{\perp}=0$
there is only state with $H_{\parallel}\times {\bf q}_{FF}
\uparrow \downarrow x$.}
\end{figure}

In the absence of parallel magnetic field the ground FF state is
two-fold degenerative due to existence of two states with opposite
directions of ${\bf q}_{FF}$ along the strip and for both
directions magnetization curves $M_x(H_{\perp})$ coincide. In Ref.
\cite{Marychev-2018} it was shown that parallel magnetic field
removes this degeneracy and makes the state with
$H_{\parallel}\times {\bf q}_{FF} \uparrow \downarrow x$ more
favorable while the other state has larger energy (it becomes
unstable at relatively low but finite $H_{\parallel}^*$). It
results to different $M_x(H_{\perp})$ for states with opposite
${\bf q}_{FF}$ at fixed $H_{\parallel}$ or vice versus. In Fig.
\ref{Fig:M(h)_par} we show this effect. Metastable state becomes
unstable at some $H_{\perp}$ and SFN strip switches to ground
state. Note that abrupt change in magnetization is not connected
with vortex entrance or exit but it occurs due to change of
direction of ${\bf \overline{q}}$.

\section{Summary}

We show that the SFN strip being in spatially modulated
(Fulde-Ferrell like) ground state has global paramagnetic response
in finite range of perpendicular magnetic fields, while at low and
large fields the response is diamagnetic. We demonstrate that the
found evolution of magnetic response with increasing of magnetic
field is accompanied by vanishing of the width averaged
longitudinal phase gradient $\overline{q_z}$ which is equal to
$q_{FF}$ at zero magnetic field. We argue that both paramagnetic
response and vanishing of $\overline{q_z}$ are related and they
are connected with peculiar dependence of sheet superconducting
current density on supervelocity (phase gradient) in FF state.

In relatively narrow SFN strip with width $w < w_c \sim
2q^{-1}_{FF}$, the transition from the state with $\overline{q_z}
\neq 0$ to state with $\overline{q_z} =0$ at field
$H_{\perp}=H_{II}$ is of second order and it occurs uniformly
along the strip (except its ends). At this field the paramagnetic
response is maximal. In wider strip ($w>w_c$) this transition is
accompanied by appearance of spatial modulation of both phase and
magnitude of the superconducting order parameter across the width
which leads to quantitative modification of the magnetic response.
Calculations in framework of Ginzburg-Landau model show that the
transition of the FF strip with width $w \gtrsim w_c$ and $\xi
\gtrsim q^{-1}_{FF}$ to state with $\overline{q_z}=0$ starts from
appearance of vortex-antivortex pairs near the ends of the strip
and ends up by formation of vortex-antivortex chain before the
first order transition occurs at $H_{\perp}=H_{II}$. At fields
$H_{\perp} \gg H_{II}$ both narrow and wide FF strips behave as
ordinary superconducting strip - they have diamagnetic response
and number of vortices increases with increase of $H_{\perp}$.

Parallel magnetic field removes the degeneracy, connected with two
directions of ${\bf q}_{FF}$ along the SFN strip. It results to
different magnetization curves $M_x(H_{\perp})$ depending on
parallel or antiparallel orientation of vector $H_{\parallel}
\times {\bf q}_{FF}$ and normal vector to surface of SFN strip.

Using parameters of NbN as S layer ($\rho_S=200 \mu \Omega \cdot
\text{cm}$, $D_S=0.5 \text{cm}^2/\text{s}$, $T_{c0}=10 \text{K}$)
and Au as N layer ($\rho_N=2 \mu \Omega \cdot \text{cm}$) we can
estimate geometrical parameters of SFN strip and value of
paramagnetic response (any ferromagnetic material could be used as
a F layer if it stays ferromagnetic when its thickness is about of
$\xi_F=(\hbar D_F/E_{ex})^{1/2}$ - for example alloy CuNi
\cite{Yamashita_2017}). For chosen materials $\xi_c=6.4 nm$ and
$M_0= 8 T$. From Fig. 2 it follows that for SFN strip with $w=20
\xi_c \sim 130 nm$ the maximal positive $4\pi M_x$ is of order of
a few tenth of Gauss. Therefore, to see the predicted effect the
array of SFN strips should be used and SQUID magnetometer to
measure their magnetization curves. We do not believe that
vortex-antivortex chain may exist in SFN strip because
$q^{-1}_{FF} \gg \xi_c\sim \xi$ for this system.

When we calculate $M_x(H_{\perp})$ we assume that magnetization of
F layer is not changed. In reality it may vary and it may give
additional contribution to magnetic response. One way to solve
this problem is to measure $M_x(H{\perp})$ above and below $T_c$
and than compare them. Second solution is to choose magnetic
material with in-plane magnetization having no or small number of
domains. To decrease $H_{\perp}$ one can take wide FF strip. For
example in \cite{Yamashita_2017} it was found well pronounced
$0-\pi$ transition in planar NbN/CuNi/NbN Josephson junction with
lateral size $10 \mu m \times 10\mu m$. This result says that at
least on this scale CuNi is homogenous enough. For FF strip based
on NbN/CuNi/Au with width $1 \mu m$ and $\xi_c=6.3$ nm we have
$H_s \sim 50 Oe$ which is small enough.

Authors acknowledge support from Foundation for the Advancement of
Theoretical Physics and Mathematics "Basis" (grant 18-1-2-64-2),
Russian Foundation for Basic Research (project number 19-31-51019)
and Russian State Contract No. 0035-2019-0021.

\end{document}